\documentclass[prb]{revtex4}
\usepackage{bm}
\usepackage{hyperref}

\newcommand{\be}{\begin{equation}}
\newcommand{\ee}{\end{equation}}
\newcommand{\lb}{\label}
\newcommand{\mb}{\mathbf}
\newcommand{\p}{\partial}
\newcommand{\nb} {\bm{\nabla}}

\begin{document}

\title{Can we derive the Lorentz force from Maxwell's equations?}
\author{Valery P. Dmitriyev}
\affiliation{Lomonosov University\\
P.O.Box 160, Moscow 117574, Russia} \email{dmitr@cc.nifhi.ac.ru}
\date{10 June 2002}

\begin{abstract}
The Lorentz force can be obtained from Maxwell's equations in the
Coulomb gauge provided that we assume that the electric portion of
the force acted on a charge is known, and the magnetic component
is perpendicular to the velocity of motion of the  charged
particle.
\end{abstract}

\maketitle

Strictly speaking, the Lorentz force can not be derived merely
from Maxwell's equations. To find it, additional postulates are
needed. As you will see below, these postulates are too strong in
order to view the procedure as a derivation. However, the job is
not useless. For, it helps us to comprehend the structure of
classical electrodynamics.

 We proceed from the general form of
Maxwell's equations \be \frac {1}{c} \frac{\p \mb{A}}{\p t} +
\mb{E} + \nb \varphi =0 \lb{1}\ee \be \frac{\p \mb{E}}{\p t} - c
\bm{\nabla}\times (\bm{\nabla} \times \mb{A}) + 4 \pi \rho\mb{v} =
0 \lb{2} \ee \be \bm{\nabla}\cdot\mb{E} = 4\pi\rho\lb{3}\ee \be
\frac{\p \rho}{\p t} + \bm{\nabla}\cdot(\rho\mb{v}) = 0\lb{4}\ee
which will be taken in the Coulomb gauge \be
\bm{\nabla}\cdot\mb{A} = 0\lb{5}\ee The system (1)-(5) is not
complete. For, it includes an uncoupled function $\mb{v}(\mb{x},
t)$. From the physical point of view the system of Maxwell's
equations describes only kinematics of motion of an electric
charge. In order to close it up, we must supplement (4) with a
dynamic equation \be \rho \frac{d v_i}{dt} = \frac{\p
\sigma_{ik}}{\p x_k} + \rho f_i\lb{6}\ee where a stress tensor
$\sigma_{ik}$ and the term $\mb{f}$ of an external force should be
defined. The portions of $\mb{f}$ are found from the same
Maxwell's equations. Thus, the problem can be posed as follows: to
define the minimal set of additional assumptions and, using them,
to extract $\mb{f}$ from (1)-(5). Before introducing new
assertions we will do some preparatory work for the second step of
the problem.

From (1) and (2) the well-known integral can be obtained:
\cite{Landau} \be \frac{1}{8\pi}\frac{\p }{\p t}\int[\mb{E}^2 +
(\nb\times\mb{A})^2]d^3x + \int\rho \mb{v}\cdot\mb{E}d^3x =
0\lb{7}\ee Manipulating (1) and (2) in another manner  we may
construct the following relation: \be \frac{1}{8\pi}\frac{\p }{\p
t}\int[(\frac{\p \mb{A}}{c\p t})^2 +
(\bm{\nabla}\!\times\mb{A})^2]d^3x - \int\rho\mb{v}\cdot
\frac{\p\mb{A}}{c\p t}\, d^3x = 0\lb{8}\ee (see Appendix
\ref{Appendix}).  We have from (1)\be \mb{E}^2 = -
\mb{E}\cdot\frac {\p\mb{A}}{c\p t} - \mb{E}\cdot\nb\varphi =
(\frac{\p\mb{A}}{c\p t})^2 + \nb\varphi\cdot\frac{\p\mb{A}}{c\p t}
- \mb{E}\cdot\nb\varphi\lb{9}\ee Substitute (\ref{9}) into (7)
taking integrals by parts and using (\ref{3}) and (5): \be
\frac{1}{8\pi}\frac{\p}{\p t}\int[(\frac{\p\mb{A}}{c\p t})^2 +
4\pi\rho\varphi + (\nb\times\mb{A})^2]d^3x + \int\rho
\mb{v}\cdot\mb{E}d^3x = 0\lb{10}\ee Subtract (\ref{8}) from (10):
\be \frac{1}{2}\frac{\p}{\p t}\int\rho\varphi d^3x + \int\rho
\mb{v}\cdot \frac{\p\mb{A}}{c\p t}d^3x + \int\rho
\mb{v}\cdot\mb{E}d^3x = 0\lb{11}\ee

Next, we will consider the system of two point electric charges at
$\mb{x_1}(t)$ and $\mb{x_2}(t)$: \be \rho (\mb{x}, t) = q_1 \delta
(\mb{x} - \mb{x_1}) + q_2 \delta (\mb{x} - \mb{x_2})\lb{12}\ee \be
\mb{v}(\mb{x}, t) = \mb{v_1}(t) I(\mb{x} - \mb{x_1}) +
\mb{v_2}(t)I(\mb{x} - \mb{x_2})\lb{13}\ee
\[\mb{v_1} = \frac{d \mb{x_1}}{d t} ,\quad\quad\quad\mb{v_2} = \frac{d \mb{x_2}}{d t}\]
where $\delta (\mb{x})$ is the Dirac delta-function and
$I(\mb{x})$ the indicator function. Substituting (12) and (13)
into (11) gives \be \frac{1}{2}(q_1 + q_2)\frac{\p \varphi}{\p t}
+
\int[(\delta(\mb{x}-\mb{x_1})q_1\mb{v_1}+\delta(\mb{x}-\mb{x_2})q_2\mb{v_2})]\cdot
\frac{\p\mb{A}}{c\p t}d^3x + (q_1\mb{v_1}+q_2\mb{v_2})\cdot\mb{E}
= 0\lb{14}\ee Remark that in Maxwell's equations the fields are
the functions of $\mb{x}$ and $t$. After the above integration
over the space coordinate they become functions of $\mb{x_1}(t)$
and $\mb{x_2}(t)$. Let us extract from (14) cross terms. We have
from Maxwell's equations for the electrostatic potential: \be
\varphi = \varphi_1 + \varphi_2\lb{15}\ee\vspace{-10 pt} \be
\varphi_1 = q_1 \phi(|\mb{x_2}-\mb{x_1}|)\lb{16}\ee\vspace{-10 pt}
\be \varphi_2 = q_2 \phi(|\mb{x_1}-\mb{x_2}|)\lb{17}\ee where a
function $\phi$ is written in variables after the integration.
That gives for cross terms in the first term of (14):\be
\frac{1}{2}(q_1\varphi_2 + q_2\varphi_1) = q_1\varphi_2 =
q_2\varphi_1\lb{18}\ee We have from Maxwell's equations for the
magnetic vector-potential: \be \mb{A} = \mb{A_1} + \mb{A_2}
\lb{19}\ee \vspace{-10 pt}\be \mb{A_1} = q_1\mb{v_1}\alpha(|\mb{x}
- \mb{x_1}|)\lb{20}\ee\vspace{-10 pt} \be \mb{A_2} =
q_2\mb{v_2}\alpha(|\mb{x}-\mb{x_2}|)\lb{21}\ee
 where a function
$\alpha$ is written in variables before the integration. Then we
have for cross terms in the second term of (14):
\[\frac{\p\mb{A_1}}{\p t} = q_1\alpha\frac{\p \mb{v_1}}{\p t} + q_1
\mb{v_1}\frac{\p\mb{x_1}}{\p
t}\cdot\frac{\p\alpha(|\mb{x}-\mb{x_1}|)}{\p
\mb{x_1}}\]\[\frac{\p\mb{A_2}}{\p t} = q_2\alpha\frac{\p
\mb{v_2}}{\p t} + q_2 \mb{v_2}\frac{\p\mb{x_2}}{\p
t}\cdot\frac{\p\alpha(|\mb{x}-\mb{x_2}|)}{\p\mb{x_2}}\]
\be\int\delta(\mb{x}-\mb{x_1})q_1\mb{v_1}\cdot\frac{\p\mb{A_2}}{\p
t}d^3x = q_1q_2[\alpha\mb{v_1}\cdot\frac{\p\mb{v_2}}{\p t} +
(\mb{v_1}\cdot\mb{v_2})\frac{\p\mb{x_2}}{\p
t}\cdot\frac{\p\alpha(|\mb{x_1}-\mb{x_2}|)}{\p
\mb{x_2}}]\lb{22}\ee
\be\int\delta(\mb{x}-\mb{x_2})q_2\mb{v_2}\cdot\frac{\p\mb{A_1}}{\p
t}d^3x = q_1q_2[\alpha\mb{v_2}\cdot\frac{\p\mb{v_1}}{\p t} +
(\mb{v_2}\cdot\mb{v_1})\frac{\p\mb{x_1}}{\p
t}\cdot\frac{\p\alpha(|\mb{x_2}-\mb{x_1}|)}{\p
\mb{x_1}}]\lb{23}\ee Summing up (22) and (23) we get the cross
terms of the second term in (\ref{14}): \[
\frac{1}{c}q_1q_2[\alpha \frac{\p (\mb{v_1}\cdot\mb{v_2})}{\p
t}+(\mb{v_1}\cdot\mb{v_2})\frac{\p \alpha}{\p
t}]=\frac{1}{c}q_1q_2\frac{ \p}{\p
t}(\alpha\mb{v_1}\cdot\mb{v_2})\] \be=\frac{1}{c}q_1\frac{ \p}{\p
t}(\mb{v_1}\cdot\mb{A_2})=\frac{1}{c}q_2\frac{ \p}{\p
t}(\mb{v_2}\cdot\mb{A_1})\lb{24}\ee We have for the electric field
\be \mb{E} = \mb{E_1} + \mb{E_2}\lb{25}\ee Then cross terms in the
third term of (\ref{14}) will be \be q_1 \mb{v_1}\cdot\mb{E_2} +
q_2 \mb{v_2}\cdot\mb{E_1}\lb{26}\ee Gathering (\ref{18}), (24) and
(26) gives for (14):  \be \frac{\p}{\p t}(q_1\varphi_2 +
\frac{1}{c}q_1\mb{v_1}\cdot\mb{A_2} + \varepsilon_0) + q_1
\mb{v_1}\cdot\mb{E_2} + q_2 \mb{v_2}\cdot\mb{E_1} + w_0 =
0\lb{27}\ee where $\varepsilon_0$ and $w_0$ are self-interaction
terms. Expression (27) was obtained from Maxwell's equations and
it is a key relation for further calculations.

Now, some assumptions will be done concerning the form of the
force term $\mb{f}$ in (\ref{6}). We postulate for the equation of
motion of a point charge \be m_1 \frac{d \mb{v_1}}{d t} = q_1
\mb{E_2} + \mb{v_1}\times (...)\lb{28}\ee The last term in (24)
means simply that the magnetic force is perpendicular to the
velocity of motion. Multiply (28) by $\mb{v_1}$: \be \frac {d}{d
t}(\frac{1}{2}m_1 \mb{v_1}^2) = q_1
\mb{v_1}\cdot\mb{E_2}\lb{29}\ee Substituting (29) into (27) we get
\be \frac{\p}{\p t}(\frac{1}{2}m_1 \mb{v_1}^2 + \frac{1}{2}m_2
\mb{v_2}^2 + q_1\varphi_2 + \frac{1}{c}q_1\mb{v_1}\cdot\mb{A_2} +
\varepsilon_0) + w_0 = 0\lb{30}\ee Expression (30) enables us to
construct the interaction Lagrangian \be L = \frac{1}{2}m_1
\mb{v_1}^2 + \frac{1}{2}m_2 \mb{v_2}^2 - q_1\varphi_2 +
\frac{1}{c}q_1\mb{v_1}\cdot\mb{A_2}\lb{31}\ee We get from (31) the
exact form of (28) for a first charge moving in the field of a
second charge \be m_1 \frac{d \mb{v_1}}{d t} = q_1 \mb{E_2} +
\frac{1}{c}q_1\mb{v_1}\times (\nb\times\mb{A_2})\lb{32}\ee

\appendix {}

\section{}\lb{Appendix}

In derivation of the integrals (\ref{7}) and (8) we proceed from
Maxwell's equations (\ref{1}) and (2).

To obtain (\ref{7}) we take the curl of (\ref{1}): \be
\frac{\p}{\p t} \nb\times\mb{A} + c\nb\times\mb{E} = 0\lb{A1}\ee
Multiply (\ref{2}) by $\mb{E}$ and (A1) by $\nb\times\mb{A}$.
Summing up the results we get \be \frac{1}{2}\frac{\p}{\p
t}\mb{E}^2 - c \mb{E}\cdot\nb\times(\nb\times\mb{A}) +
4\pi\rho\mb{v}\cdot\mb{E} + \frac{1}{2}\frac{\p}{\p
t}(\nb\times\mb{A})^2 + c(\nb\times\mb{E})\cdot(\nb\times\mb{A}) =
0\lb{A2}\ee Integrate (A2) over the whole space and take the
second integral by parts supposing that the fields are vanishing
at infinity. Then the respective integrals obtained from the
second and fifth terms of (A2) cancel each other. Thus we come to
(\ref{7}) sought for.

In order to derive (\ref{8}) we firstly operate (\ref{1}) with
$\p_t$: \be \frac{\p^2\mb{A}}{c\p t^2} + \frac{\p\mb{E}}{\p t} +
\bm{\nabla}\frac{\p\varphi}{\p t}
 = 0\lb{A3}\ee Then exclude $\p\mb{E}/\p t$ from (A3) and (\ref{2}):
\be \frac{\p^2\mb{A}}{c\p t^2} + c\nb\times(\nb\times\mb{A}) +
\bm{\nabla}\frac{\p\varphi}{\p t}
 = 4\pi\rho\mb{v}\lb{A4}\ee Multiply (A4) by $\p\mb{A}/\p t$:
\be \frac{1}{2}\frac{\p}{c\p t}(\frac{\p\mb{A}}{\p t})^2 +
c\frac{\p\mb{A}}{\p t}\cdot\nb\times(\nb\times\mb{A}) +
\frac{\p\mb{A}}{\p t}\cdot\bm{\nabla}\frac{\p\varphi}{\p t}
 = 4\pi\rho\mb{v}\cdot\frac{\p\mb{A}}{\p t}\lb{A5}\ee
Integrate (A5) over the whole space. Take the intergals of the
second and third terms by parts. The third integral vanishes due
to (\ref{5}). Thus we arrive at the relation (\ref{8}) sought for.

\end{document}